%% file: apssamp.tex
\documentclass[prb,twocolumn,amsmath,amssymb,superscriptaddress]{revtex4-1}

\usepackage[normalem]{ulem}
\usepackage{graphicx}
\usepackage{todonotes} 
\usepackage{dcolumn}
\usepackage{tabularx}
\usepackage{bm}
\usepackage{hyperref}
\usepackage{textcomp}
\hypersetup{
    colorlinks=true,
    linkcolor=blue,
    citecolor = blue,
    filecolor=magenta,      
    urlcolor=blue,
    pdftitle={Broadband CPW-based impedance-transformed Josephson parametric amplifier},
    pdfpagemode=FullScreen,
    }

\usepackage{multirow}
\usepackage{nicefrac}
\usepackage{qcircuit}
\usepackage{physics}

\begin{document}

\preprint{APS/123-QED}

\title{Broadband CPW-based impedance-transformed Josephson parametric amplifier}

\author{Bingcheng Qing}
\email{bc.qing@berkeley.edu}
\affiliation{Department of Physics, University of California, Berkeley, California 94720, USA}

\author{Long B. Nguyen}
\email{longbnguyen@berkeley.edu}
\affiliation{Department of Physics, University of California, Berkeley, California 94720, USA}
\affiliation{Computational Research Division, Lawrence Berkeley National Laboratory, Berkeley, California 94720, USA}

\author{Xinyu Liu}
\affiliation{Anyon Technologies, Emeryville, California 94662, USA}

\author{Hengjiang Ren}
\affiliation{Anyon Technologies, Emeryville, California 94662, USA}

\author{William P. Livingston}
\altaffiliation{Current address: Google Quantum AI, Mountain View, CA 94043, USA}
\affiliation{Department of Physics, University of California, Berkeley, California 94720, USA}
\affiliation{Computational Research Division, Lawrence Berkeley National Laboratory, Berkeley, California 94720, USA}

\author{Noah Goss}
\affiliation{Department of Physics, University of California, Berkeley, California 94720, USA}
\affiliation{Computational Research Division, Lawrence Berkeley National Laboratory, Berkeley, California 94720, USA}

\author{\\Ahmed Hajr}
\affiliation{Department of Physics, University of California, Berkeley, California 94720, USA}

\author{Trevor Chistolini}
\affiliation{Department of Physics, University of California, Berkeley, California 94720, USA}

\author{Zahra Pedramrazi}
\affiliation{Department of Physics, University of California, Berkeley, California 94720, USA}
\affiliation{Computational Research Division, Lawrence Berkeley National Laboratory, Berkeley, California 94720, USA}

\author{David I. Santiago}
\affiliation{Department of Physics, University of California, Berkeley, California 94720, USA}
\affiliation{Computational Research Division, Lawrence Berkeley National Laboratory, Berkeley, California 94720, USA}

\author{Jie Luo}
\affiliation{Anyon Technologies, Emeryville, California 94662, USA}

\author{Irfan Siddiqi}
\affiliation{Department of Physics, University of California, Berkeley, California 94720, USA}
\affiliation{Computational Research Division, Lawrence Berkeley National Laboratory, Berkeley, California 94720, USA}

\begin{abstract}
    \input{sections/0_abstract}
\end{abstract}

\maketitle

\input{sections/1_intro}

\input{sections/2_concept}

\input{sections/4_performance}

\input{sections/6_summary}

\noindent \textbf{Acknowledgments}

\noindent We thank Larry Chen and Kan-Heng Lee for providing the transmon sample. This work was supported by the Quantum Testbed Program of the Advanced Scientific Computing Research Division, Office of Science of the U.S. Department of Energy under Contract No. DE-AC02-05CH11231.
\\

\input{sections/author_declarations}


\bibliography{apssamp}

\clearpage

\pagebreak
\widetext
\begin{center}
\textbf{\large Supplementary Materials}
\end{center}
\setcounter{equation}{0}
\setcounter{figure}{0}
\setcounter{table}{0}
\setcounter{page}{1}
\makeatletter
\renewcommand{\theequation}{S\arabic{equation}}
\renewcommand{\thefigure}{S\arabic{figure}}

\section{Impedance Transformer Parameters}

\begin{figure}[htbp]
    \centering
    \includegraphics[width=0.65\textwidth]{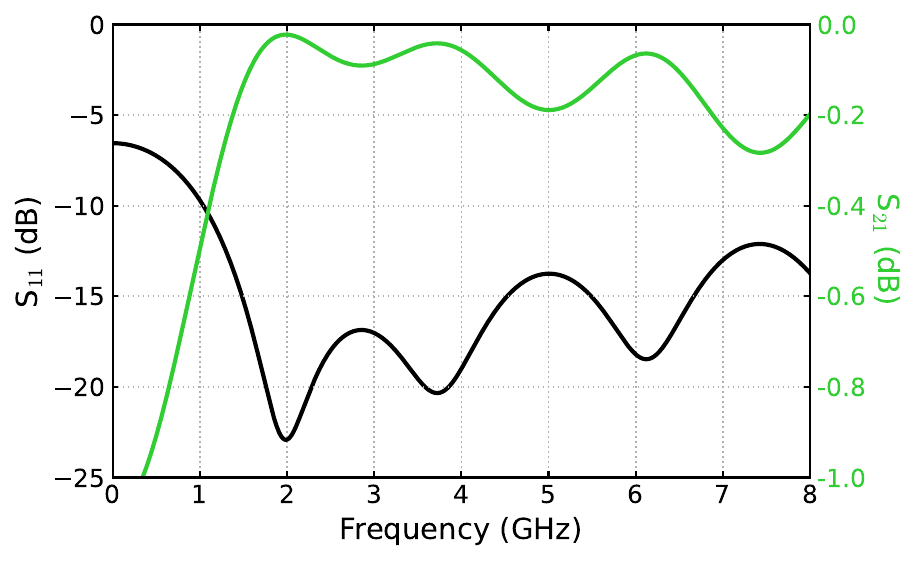}
    \caption{S-matrix simulation result of the horn-like CPW line.}
    \label{fig:S_matrix}
\end{figure}

To construct the impedance transformer, we tailor a CPW with its gap fixed at 3 $\mathrm{\mu m}$ and its center trace width tapered from 7.8~$\mathrm{\mu m}$ to 218~$\mathrm{\mu m}$, forming a horn-like shape. The characteristic impedance along this horn-like CPW line follows the Klopfenstein taper profile~\cite{pozar2011microwave}, which has a high-pass feature that suppresses reflection above a cutoff frequency. The impedance transformer here is designed to have a maximum reflection of -10 dB above 2 GHz. We note that it can be shortened by half if one increases the cutoff frequency to 4 GHz, which further reduces the complexity and footprint size of the CIMPA. The Klopfenstein profile is calculated using MATLAB, and the CPW characteristic impedance is calculated using TX-Line, a convenient industrial-level transmission line calculator~\cite{simons2004coplanar}. To verify the property of the impedance transformer, we simulate the transmission and reflection coefficients (S$_{21}$ and S$_{11}$, respectively) of the horn-like CPW line with ANSYS HFSS. The result in Fig.~\ref{fig:S_matrix} shows low reflection (below -10 dB) and high transmission (insertion loss below 0.3 dB) above the intended 2-GHz cutoff frequency.

\section{Device fabrication}

The device is fabricated on an intrinsic high-resistivity Si wafer with resistivity $\rho>$ 10~$\mathrm{k\Omega} \text{-}\mathrm{cm}$. After sputtering a 100-nm-Al film, the horn-like CPW's features are defined using photo-lithography and reactive-ion etching. Notably, the fabrication process is compatible with the photo-lithography technique thanks to the 3-$\mathrm{\mu m}$ gap. Then, the capacitor is fabricated using the lift-off e-beam lithography technique followed by e-beam evaporation of aluminum oxide and aluminum electrodes. We note that the size of the capacitor is approximately 39$\times$39 $\mathrm{\mu m}^2$, which can also be made using photo-lithography. The SQUID structure is defined using e-beam lithography and double-angle e-beam evaporation of aluminum films following the standard Dolan bridge technique~\cite{dolan1977offset}. The galvanic contact between the SQUID and the capacitor electrodes is formed using the Argon ion-milling band-aid process~\cite{potts2001cmos}. Finally, a layer of methyl methacrylate (MMA) resist is used to protect the wafer before dicing. The chips are diced and then cleaned using N-methylpyrrolidone (NMP) before being packaged into a copper box.

\section{Device measurement setup and calibration}
\begin{figure}[t!]
    \centering
    \includegraphics[width=0.9\textwidth]{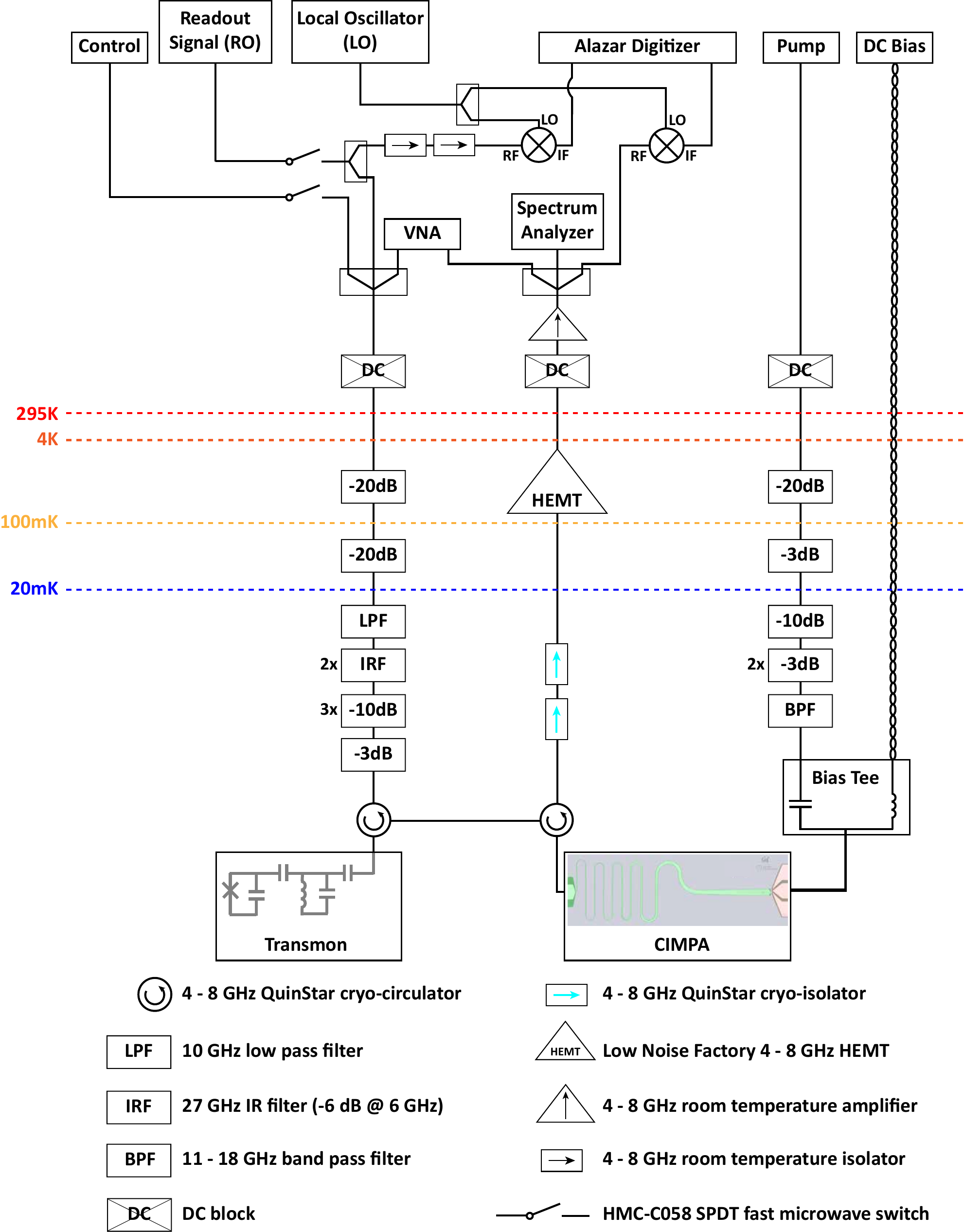}
    \caption{The wiring diagram of the fridge}
    \label{fig:wiring}
\end{figure}

The CIMPA is integrated into a transmon qubit measurement setup and characterized inside a dilution fridge. The detailed fridge wiring diagram is shown in Fig. \ref{fig:wiring}. 

A Rohde \& Schwarz SGT100A vector RF source (Control) is used to generate qubit control pulses. An Agilent Technologies E8257D signal generator (RO) generates the readout signal at the readout resonator frequency. By mixing the readout signal with a Hewlett Packard 8671B local oscillator (LO), the readout signal information is converted to an intermediate frequency of 50 MHz for the AlazarTech digitizer (Alazar) to record the signal. The readout signal and control signal are pulse-controlled using HMC-C058 SPDT switches. A Tektronix 5014C arbitrary waveform generator generates the pulse sequences to control the qubit, readout the resonator signals, and trigger the Alazar. The DC bias current to the CIMPA is provided by a Yokogawa GS200 DC voltage/current source, and the AC pump is supplied by a Hewlett Packard 83732B signal generator. The CIMPA gain and noise performance are measured using an Agilent Technologies N9010A signal analyzer (SA) and an Agilent Technologies N5230C network analyzer (VNA). 

\begin{figure}[th]
    \centering
    \includegraphics[width=0.9\textwidth]{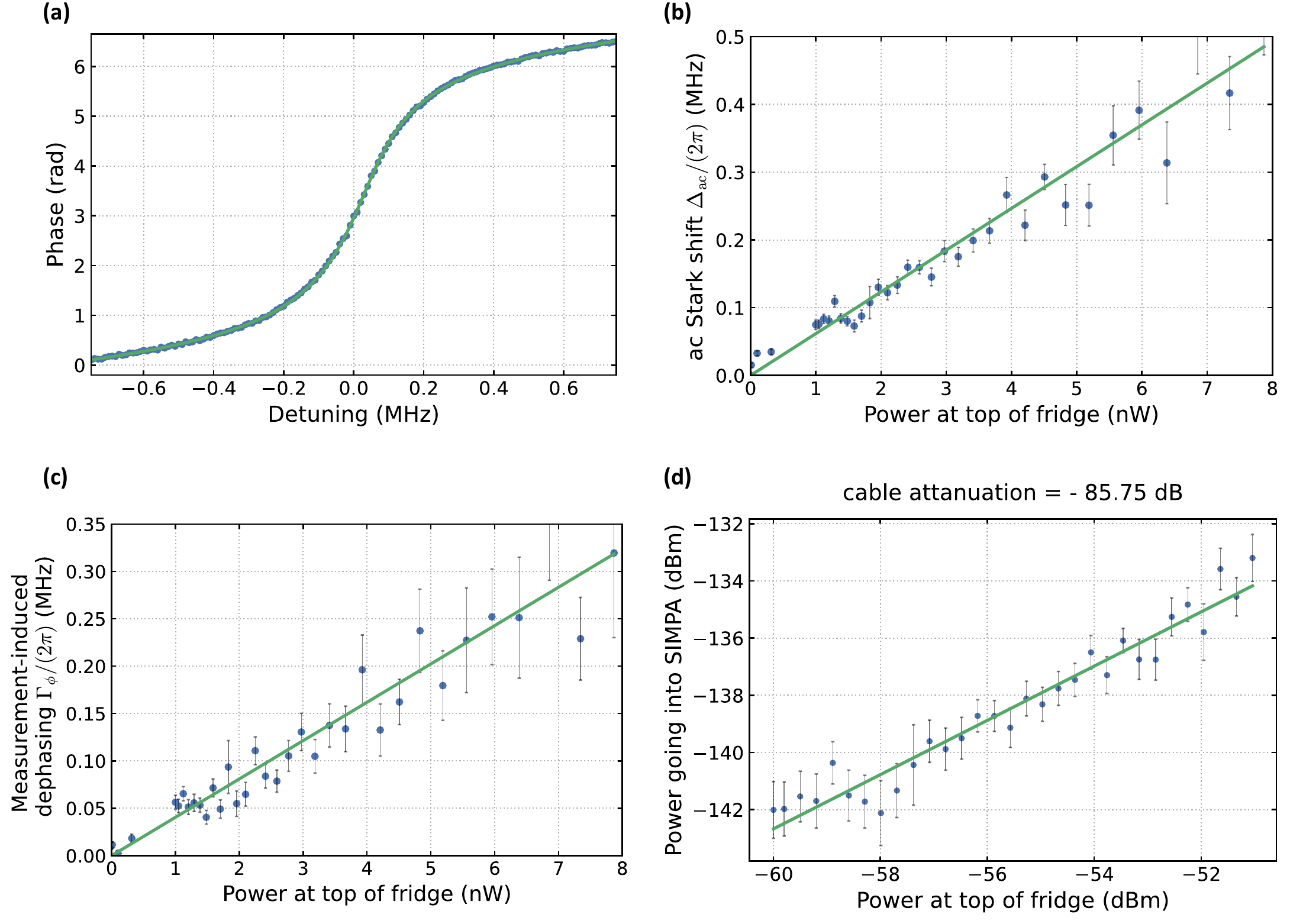}
    \caption{The calibration of the measurement setup. (a) Fitting of the readout resonator. (b) Fitting of the ac Stark shift. (c) Fitting of the measurement-induced dephasing. (d) Calibration between the power going into CIMPA and the power at the top of the fridge}
    \label{fig:calibration}
\end{figure}

To precisely measure the input power of the amplifier, we need to calibrate the attenuation along the input line and convert the power sent out of the VNA to the power fed into CIMPA. According to the method mentioned in Ref.~\cite{macklin2015near}, we can calibrate the input signal power by extracting the intra-cavity photon number occupation $\bar{n}$ inside the readout resonator. The power of the signal leaving the qubit and feeding into CIMPA is given by 
\begin{equation}
    P = \hbar \omega_r\kappa_r\bar{n},
\end{equation}
where $\omega_r$ and $\kappa_r$ are the angular frequency and linewidth of the readout resonator, respectively. The resonator linewidth and frequencies are extracted by directly fitting the resonator reflection spectrum, as shown in Fig.~\ref{fig:calibration}(a). 

The intra-cavity photon number is measured using the ac Stark shift and measurement-induced dephasing of the qubit, which can be extracted by fitting the Ramsey measurement. Considering a qubit dispersively coupled to a readout resonator with a dispersive shift $\chi$, the ac Stark shift caused by the photon population in the readout resonator is $\Delta_{ac} = 2\chi\bar{n}$, and the measurement-induced dephasing is given by $\Gamma_\phi = 8\chi^2\bar{n}/\kappa_r$\cite{gambetta2006qubit}. Utilizing the quadratic dependence of the dephasing rate and the linear dependence of the frequency shift on $\chi$, we can extract $\chi$ and $\bar{n}$. As we vary the VNA power outside the fridge, we can extract the intra-cavity photon number of the readout resonator by the Ramsey spectrum of the qubit, and then calibrate the corresponding signal power at the CIMPA. The ac Stark shift and measurement-induced dephasing at various VNA powers are shown in Fig. \ref{fig:calibration}(b, c). The extracted correspondence between the signal sent into CIMPA and the VNA power outside the fridge is shown in Fig.~\ref{fig:calibration}~(d). All extracted parameters are listed in Table~\ref{qubit_param}.

\newcolumntype{L}[1]{>{\raggedright\let\newline\\\arraybackslash\hspace{0pt}}m{#1}}
\newcolumntype{C}[1]{>{\centering\let\newline\\\arraybackslash\hspace{0pt}}m{#1}}
\newcolumntype{R}[1]{>{\raggedleft\let\newline\\\arraybackslash\hspace{0pt}}m{#1}}
\begin{table}[htbp]
    \centering
    \begin{tabular}{|C{2cm}||C{2.5cm}|C{2.5cm}|C{2.5cm}|C{2.5cm}|C{2.5cm}|}
    \hline
         Parameter&$\omega_q/(2\pi)$ (GHz)& $\omega_r/(2\pi)$ (GHz)& $\kappa_r/(2\pi)$ (kHz)& $T_1$ ~($\mathrm{\mu s}$)& $T_{2,\text{echo}}$ ~($\mathrm{\mu s}$) \\
         \hline
         Value&5.350  & 6.633  & 309  & 45 & 16 \\
         \hline
    \end{tabular}
    \caption{Qubit and resonator parameters}
    \label{qubit_param}
\end{table}

\section{Derivation of CIMPA Hamiltonian and gain}
The theoretical derivation in this section is based on Ref. \cite{eddins2017superconducting} The gain of the amplifier originates from the 3-wave mixing process enabled by the SQUID in the nonlinear LC resonator. If we DC bias the SQUID at a specific flux point and pump it using a small AC flux tone $\Phi_e = \frac{\Phi_0}{\pi}(F + \delta f\cos(\omega_p t))$, we can write down its Hamiltonian as
\begin{equation}
    \mathcal{H} = 4E_cn^2 - E_{\text{J,eff}}\left|\cos(F + \delta f\cos(\omega_p t))\right|\cos(\phi),
    \label{eq:Hamiltonian_sp}
\end{equation}
where $E_c = e^2/(2C)$ is the single electron capacitive energy, $E_\text{J,eff} = \Phi_0^2/(4\pi^2L_\text{J,eff})$ is the effective Josephson energy of the SQUID, $\Phi_0$ is the magnetic flux quantum, $n$ and $\phi$ are two conjugate variables of the circuit: Cooper-pair number on the capacitor and gauge-invariant phase across the SQUID, respectively. After the quantization of the Hamiltonian, one can write down the Hamiltonian in the rotating frame with the bosonic modes $\hat{a}$ of the CIMPA and apply rotating wave approximation,
\begin{equation}
    \hat{\mathcal{H}}/\hbar = \Delta \hat{a}^\dagger \hat{a} + \lambda (\hat{a}^{\dagger2} + \hat{a}^2) + \hat{O}(\hat{a}^4),
    \label{eq:QHamiltonian_sp}
\end{equation}
where $\lambda = \delta fE_\text{J,eff}\sqrt{\sin(F)\tan(F)E_c/(8E_\text{J,eff})}$ originates from flux modulation, $\Delta = \omega_p/2 - \omega_0$ is the detuning between the half pump frequency and the CIMPA resonance frequency, which will be assumed to be zero in the following discussion. Now we can use the input-output theory to write down the equation of motion for the CIMPA coupled to a transmission line with decay rate $\kappa$,
\begin{equation}
    \dot{\hat{a}} = \frac{i}{\hbar}[\hat{\mathcal{H}}, \hat{a}] - \frac{\kappa}{2}\hat{a} + \sqrt{\kappa}\hat{a}_\text{in} = -2i\lambda\hat{a}^\dagger - \frac{\kappa}{2}\hat{a} + \sqrt{\kappa}\hat{a}_\text{in}.
    \label{eq:InOut_sp}
\end{equation}

We can solve this equation of motion in the frequency domain by defining $\Bar{a}[\omega] = \int_{-\infty}^{\infty} \text{d}t\ e^{i\omega t} \hat{a}(t)$ as the frequency-domain bosonic annihilation operator, as shown in Eq. \ref{eq:solution_sp}. One should notice here that $\omega$ is the detuning from the CIMPA resonance frequency because we work in the rotating frame,
\begin{equation}
    \bar{a}[\omega]=\sqrt{\kappa}\left(\frac{(\kappa / 2-i \omega) \bar{a}_{\text {in }}[\omega]-2i \lambda \bar{a}_{\text {in }}^{\dagger}[-\omega]}{|\kappa / 2-i \omega|^2-4|\lambda|^2}\right).
    \label{eq:solution_sp}
\end{equation}
Finally, with the input-output boundary condition $\sqrt{\kappa}\hat{a} = \hat{a}_\text{in} + \hat{a}_\text{out}$, we can obtain the relation between the input signal tone $\hat{a}_\text{in}$ and output signal tone $\hat{a}_\text{out}$,

\begin{equation}
\begin{split}
    \bar{a}_{\text {out }}[\omega]= \left( \kappa\frac{\kappa / 2-i \omega}{|\kappa / 2-i\omega|^2-4|\lambda|^2}-1 \right) \bar{a}_{\text {in }}[\omega]\\
    +\left(\frac{-2i \kappa \lambda}{|\kappa / 2-i \omega|^2-4|\lambda|^2}\right) \bar{a}_{\text {in }}^{\dagger}[-\omega],
\end{split}
\label{eq:InOutSolution_sp}
\end{equation}
from which we can identify the gain $G(\omega)$ as the ratio between the power of output and input signals with the same frequency,
\begin{equation}
    G(\omega) = \left| \kappa\frac{\kappa / 2-i \omega}{|\kappa / 2-i\omega|^2-4|\lambda|^2}-1 \right|^2.
    \label{eq:gain_sp}
\end{equation}

\section{Dependence on pump power and signal power}

\begin{figure}[htbp]
    \centering
    \includegraphics[width=0.8\textwidth]{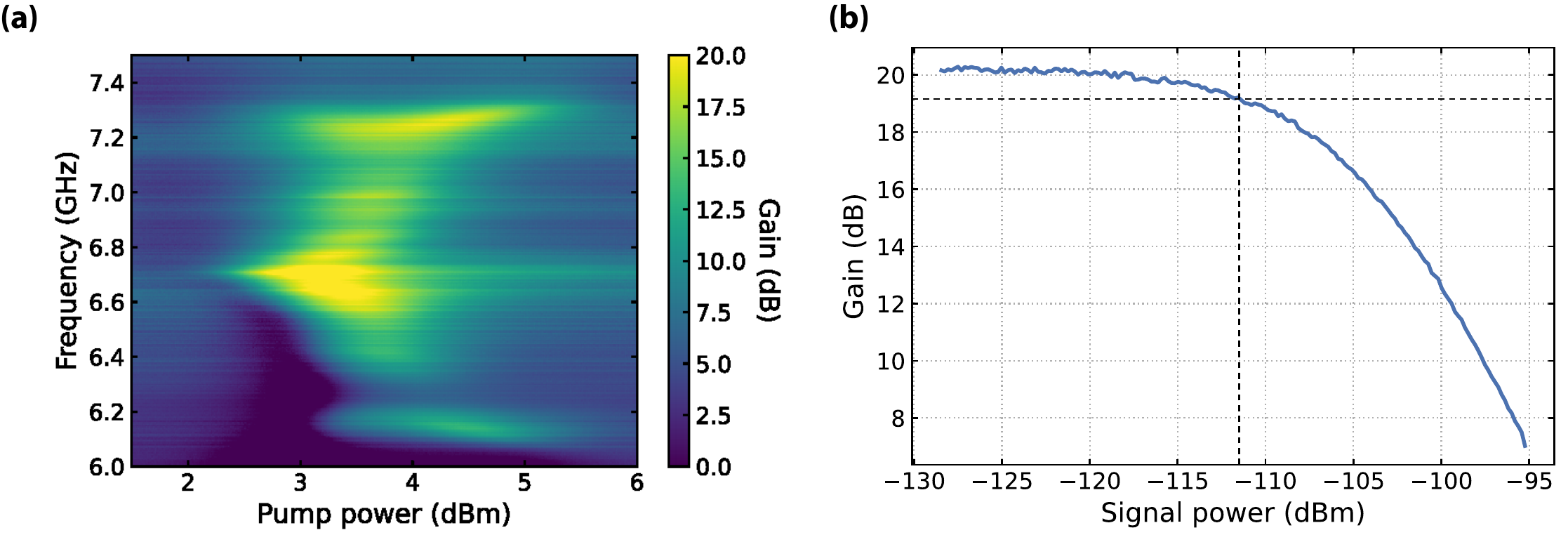}
    \caption{Gain performance with different (a) pump power and (b) signal power.}
    \label{fig:Gain_pump}
\end{figure}

CIMPA offers gain near the instability point where the denominator of the gain formula shown in Eq.~\ref{eq:gain_sp} approaches zero. Therefore, a small change in the denominator will greatly affect the gain. There are usually three factors influencing the denominator of the gain profile: pump frequency, pump power, and signal power. The pump frequency and power appear in the denominator formula, and the signal power introduces an ac Stark shift which effectively detunes the signal and decreases the gain. With a normal industrial-level signal generator, the precision of pump frequency can be as high as 1 Hz, but the pump power and signal power can fluctuate more. Therefore, it is worth examining the sensitivity of CIMPA gain performance with respect to the pump power and signal power.
Figure~\ref{fig:Gain_pump}(a) shows the gain profile at different signal frequencies while sweeping the power of the pump signal. The pump frequency is fixed at 13.41~GHz and the device is DC biased at $\Phi_e = 0.3\Phi_0$. We observe reasonable gain over around 1 dBm range of the pump power.

The sensitivity of the gain profile with respect to the signal power indicates the saturation effect. The gain profile is expected to be almost unchanged at small signal power to offer a linear gain and decrease at large signal power due to saturation. Figure~\ref{fig:Gain_pump}(b) shows the gain of a signal at 6.665 GHz as we increase the signal power. The device is operated with 13.41-GHz pump frequency and $\Phi_e = 0.3\Phi_0$ DC flux bias point. We observe an almost constant signal gain below the 1-dB compression point of -111.5 dBm, and a fast drop in gain after this value. The saturation input power is similarly extracted for different operation parameters.

\end{document}

%% file: sections/0_abstract.tex
Quantum-limited Josephson parametric amplifiers play a pivotal role in advancing the field of circuit quantum electrodynamics by enabling the fast and high-fidelity measurement of weak microwave signals. Therefore, it is necessary to develop robust parametric amplifiers with low noise, broad bandwidth, and reduced design complexity for microwave detection. However, current broadband parametric amplifiers either have degraded noise performance or rely on complex designs. Here, we present a device based on the broadband impedance-transformed Josephson parametric amplifier (IMPA) that integrates a horn-like coplanar waveguide (CPW) transmission line, which significantly decreases the design and fabrication complexity, while keeping comparable performance. The device shows an instantaneous bandwidth of 700(200) MHz for 15(20) dB gain with an average saturation power of -110 dBm and near quantum-limited added noise. The operating frequency can be tuned over 1.4 GHz using an external flux bias. We further demonstrate the negligible back-action from our device on a transmon qubit. The amplification performance and simplicity of our device promise its wide adaptation in quantum metrology, quantum communication, and quantum information processing.


%% file: sections/1_intro.tex
\vspace{1em}
\noindent\textbf{Introduction}

\noindent 
The parametric process facilitates coherent energy transfer among electromagnetic waves based on nonlinearities, enabling quantum-limited parametric amplification~\cite{caves1982quantum}. Microwave parametric amplification not only improves the readout of solid-state~\cite{blais2021circuit,burkard2023semiconductor,vijay2011observation,hatridge2013quantum,arute2019quantum} and nanomechanical~\cite{arash2015review} quantum devices, but also becomes increasingly important in emerging technologies such as hybrid quantum systems~\cite{clerk2020hybrid} and axion dark matter detection~\cite{caldwell2017dielectric,nitta2023search}. It is thus important to develop amplifiers with outstanding performance: low noise, broad bandwidth, and high saturation power. Moreover, for easy adaptation of parametric amplifiers, their design and fabrication should be robust and simple.

In superconducting circuits, parametric amplification is usually achieved by Josephson parametric amplifiers (JPAs) utilizing multi-wave mixing processes in nonlinear superconducting resonators, where Josephson junctions provide the nonlinearity without introducing dissipation~\cite{roy2016,manucharyan2007microwave}. The nonlinear resonator of the JPA is typically coupled to a fixed 50-$\Omega$ embedding environment~\cite{castellanos2007widely,yamamoto2008flux,bergeal2010phase,hatridge2011dispersive}, which prevents tuning its linewidth and saturation power independently, resulting in a trade-off between saturation power and bandwidth\cite{mutus2014strong}. Hence, its bandwidth is inherently limited to 10-50 MHz, and its saturation power is typically -120 dBm for a nominal gain of 20 dB.  

There are two strategies to overcome this limitation. In the first approach, the nonlinear resonator is unpacked into a nonlinear transmission line by thousands of Josephson junctions, forming a traveling-wave-parametric-amplifier (TWPA), in which the gain is optimized across a broad range of frequencies~\cite{esposito2021perspective}. Such a device must satisfy two conditions: (i) the impedance of the amplifier has to match the 50-$\Omega$ environmental impedance, and (ii) the phases of the pump, signal, and idler tones have to match~\cite{esposito2021perspective}. To fulfill these challenging requirements, 2000 nonlinear unit cells were constructed to form a TWPA with a 20-dB-gain bandwidth over 3 GHz and a saturation input power of -99 dBm~\cite{macklin2015near}. However, TWPA noise processes can be complex~\cite{esposito2021perspective}, and the stringent requirements on the phase matching and the large number of Josephson junctions make the design, calibration and fabrication difficult for TWPAs~\cite{ezenkova2022broadband}.


In the second approach, the environmental impedance is engineered to enhance the JPA's coupling, which leads to enhanced bandwidth and saturation power~\cite{mutus2014strong}. This has been accomplished by both simple narrowband impedance transformers~\cite{roy2015broadband,grebel2021flux} and more complex broadband impedance transformers~\cite{mutus2014strong,lu2022broadband,ranzani2022wideband}. Narrowband impedance transformers constructed using $\lambda/4$ and $\lambda/2$ coplanar-waveguides (CPWs)~\cite{roy2015broadband,grebel2021flux} enhances the JPA's bandwidth to 200~MHz with fixed operating frequency. JPAs with broadband impedance transformers following Klopfenstein~\cite{mutus2014strong,lu2022broadband} or Ruthroff~\cite{ranzani2022wideband} taper designs have instantaneous bandwidth in the range of 600-700~MHz and allow tunable operating frequencies, but their current implementations either require multi-layer fabrication with lossy dielectrics~\cite{mutus2014strong, ranzani2022wideband} or careful simulation and calibration of finely patterned interdigitated capacitors~\cite{ranzani2022wideband}. These requirements increase the complexity of the amplifiers' design, calibration, and fabrication.

\begin{figure*}[t!]
    \centering
    \includegraphics[width=0.85\textwidth]{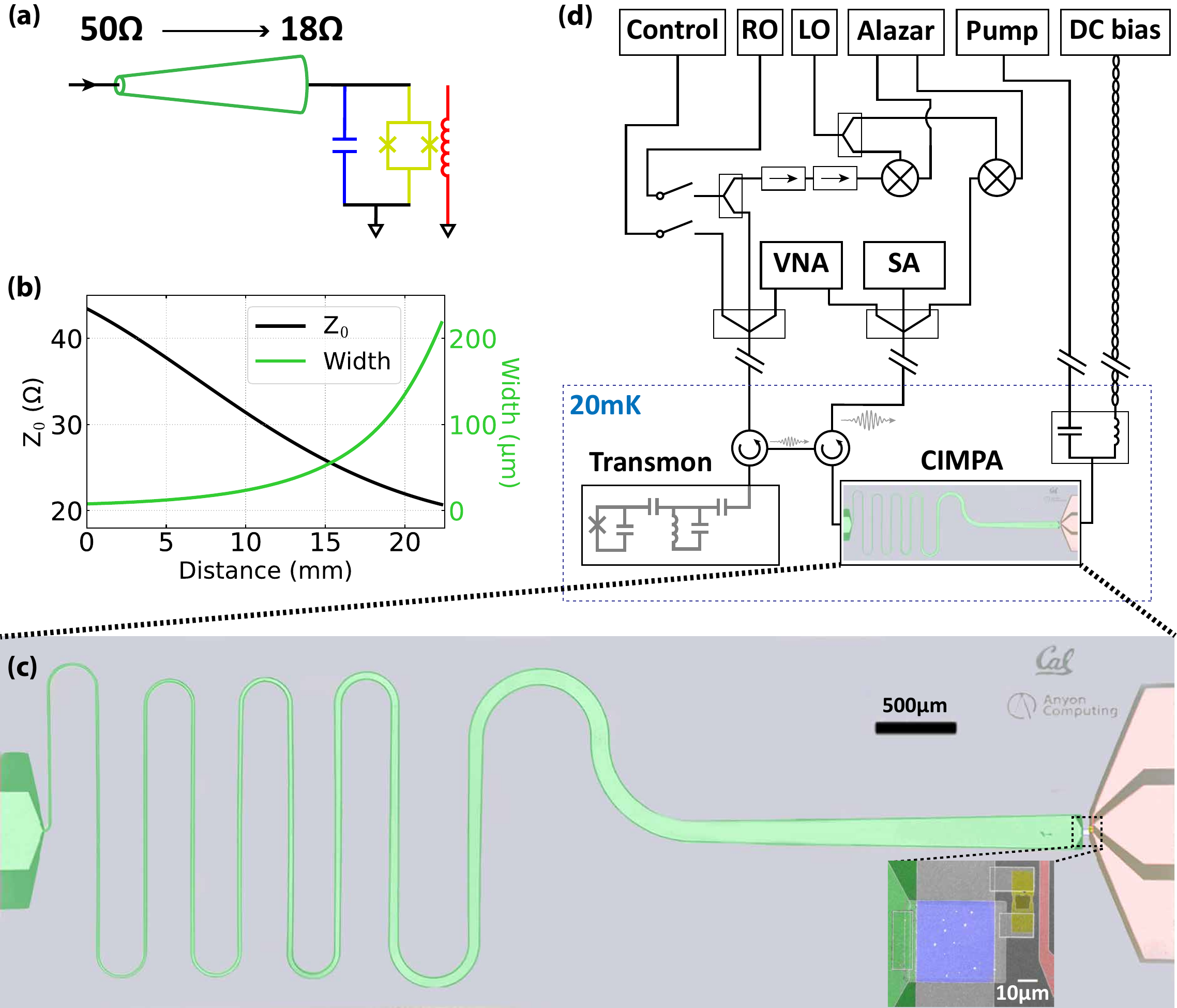}
    \caption{Experimental concept. (a) Electrical circuit depiction of the CIMPA. The nonlinear resonator on the right side is constructed by shunting a SQUID (yellow) with a capacitor (blue). The structure is galvanically coupled to a Klopfenstein taper (green) on the left, and inductively coupled to a flux-line (red) on the right. (b) Parameters of the horn-like CPW impedance transformer, showing decreasing Z$_o$ along the CPW as its center-trace width becomes wider, varying from 50 $\Omega$ at the left end to 18 $\Omega$ at the right end. (c) False-color micrograph of the prototype sample with the nonlinear resonator shown in the inset. From left to right are the horn-like CPW taper (green), the capacitor (blue) and SQUID (yellow) forming the resonator, and the on-chip flux line (red). (d) Simplified diagram of the measurement apparatus (details in Suppl. Mat.). A transmon qubit is utilized to calibrate the input power and detect possible back action from the CIMPA.}
    \label{fig:device}
\end{figure*}

In this Letter, we present a CPW-based broadband impedance-transformed parametric amplifier (CIMPA) design and showcase its performance. By reducing the technical requirements of the Klopfenstein taper, we can implement it with a single horn-like CPW, which keeps the broadband behavior while maintaining the simplicity of the design and fabrication. The horn-like CPW structure allows us to leverage streamlined RF design tools~\cite{simons2004coplanar} and minimize the complexity of simulating and fabricating the impedance transformer. Notably, the CIMPA performs as well as the other IMPAs despite such simplicity. The amplifier displays an instantaneous bandwidth of 700(200) MHz for 15(20) dB gain, a 1.4 GHz flux-tunable bandwidth, a saturation input power of approximately -110 dBm, and no significant back action on the qubit.



%% file: sections/2_concept.tex
\vspace{1em}
\noindent\textbf{Device concept}

\noindent The CIMPA is constructed from three basic building blocks, as depicted by the circuit diagram in Fig.~\ref{fig:device}(a). The core component of the device is a nonlinear LC resonator that facilitates the necessary parametric process through which amplification occurs. It consists of a superconducting quantum interference device (SQUID, yellow) and a capacitor (blue) in parallel. On the right-hand side, it is inductively coupled to a flux line (red) that allows in-situ tuning of the resonator frequency and RF flux pumping~\cite{grebel2021flux, yamamoto2008flux}. On the left-hand side, it is galvanically connected to an impedance taper (green) that is central to improving its bandwidth and dynamic range. 

The CIMPA operates as follows. When an RF microwave pump is applied to the nonlinear resonator via the flux line at approximately twice its resonant frequency $\omega_p\approx 2\omega_0$, we can apply the rotating wave approximation (RWA) in the frame rotating at the signal frequency $\omega_s\approx \omega_0$ and write the dressed Hamiltonian as~\cite{eddins2017superconducting}
    \begin{equation}
        \hat{\mathcal{H}}/\hbar=\Delta \hat{a}^\dagger \hat{a} + \lambda (\hat{a}^{\dagger2} + \hat{a}^2) + \hat{O}(\hat{a}^4),
    \label{eq:QHamiltonian}
    \end{equation}
where $\hat{a} (\hat{a}^\dagger)$ is the bosonic annihilation (creation) operator associated with the resonator, $\lambda$ originates from the modulation of the external flux by the pump, and $\Delta=\omega_p/2 - \omega_0\sim 0$. If the circuit is coupled to the environment with strength $\kappa$, its input-output equation of motion is given by
    \begin{equation}
    \dot{\hat{a}} = \frac{i}{\hbar}[\hat{\mathcal{H}}, \hat{a}] - \frac{\kappa}{2}\hat{a} + \sqrt{\kappa}\hat{a}_\text{in} = -2i\lambda\hat{a}^\dagger - \frac{\kappa}{2}\hat{a} + \sqrt{\kappa}\hat{a}_\text{in}
    \label{eq:InOut},
    \end{equation}
where $\hat{a}_\mathrm{in}$ and $\hat{a}_\mathrm{out}$ are the bosonic operators associated with incoming and outgoing signals. We then proceed to define $\Bar{a}[\omega] = \int_{-\infty}^{\infty} \text{d}t\ e^{i\omega t} \hat{a}(t)$ 
as the frequency-domain bosonic annihilation operator, and use the boundary condition $\sqrt{\kappa}\hat{a} = \hat{a}_\text{in} + \hat{a}_\text{out}$ to obtain 
    \begin{equation}
    \begin{split}
        \bar{a}_{\text {out }}[\omega]= \left[ \kappa\frac{\kappa / 2-i \omega}{(\kappa / 2-i\omega)^2-4|\lambda|^2}-1 \right] \bar{a}_{\text {in }}[\omega]\\
        +\left[\frac{-2i \kappa \lambda}{(\kappa / 2-i \omega)^2-4|\lambda|^2}\right] \bar{a}_{\text {in }}^{\dagger}[-\omega],
    \end{split}
    \label{eq:InOutSolution}
    \end{equation}
from which the gain is define, $G(\omega)=|\bar{a}_{\text {out }}/\bar{a}_{\text {in}}|^2$.

For a suitable value of $\lambda$, the denominator in Eq.~\ref{eq:InOutSolution} approaches zero, resulting in large signal gain, which remains at finite values in practice because of higher-order terms that are not included in Eq.~\ref{eq:QHamiltonian}.~\cite{boutin2017effect} In addition, large signal power may lead to an ac Stark shift and pump power depletion, thereby decreasing the effective gain~\cite{boutin2017effect}. The incoming signal power associated with a gain drop of 1 dB is defined as the saturation input power, or alternatively, the 1-dB compression point.

Importantly, Eq.~\ref{eq:InOutSolution} shows that the amplifier's bandwidth is determined by the coupling strength $\kappa$, and thereby can be improved by enhancing the coupling between the resonator and the signal. For the circuit shown in Fig.~\ref{fig:device}(a), this coupling rate is given as $\kappa = (CZ_0)^{-1}$, while the saturation power is $P_\mathrm{sat}\sim C/Z_0$\cite{mutus2014strong}, where $Z_0$ is the characteristic impedance of the coupled environment. Thus, by engineering $Z_0$, one can increase both the bandwidth and the dynamic range of the amplifier. Since the signals must come through a nominal 50-$\Omega$ transmission line before and after interacting with the amplifier, an impedance transformer is necessary for enhanced coupling, shown on the left-hand side of Fig.~\ref{fig:device}(a).

Previously, this idea was realized using a Klopfenstein taper with a characteristic impedance ranging from $50~\Omega$ at one end to $15~\Omega$ at the other end, and a maximum return loss of -20 dB above the cut-off frequency~\cite{mutus2014strong,lu2022broadband}. These stringent requirements may lead to sophisticated designs that are associated with tedious simulation procedures and complex nanofabrication processes. Alternatively, a combination of quarter-wavelength and half-wavelength CPW resonators can be used~\cite{roy2015broadband}, with the setback of having a narrower bandwidth, thereby negating the flux-tunable advantage of the SQUID. 

Our design unifies the advantages of these two approaches by relaxing the condition on maximum return loss to -10 dB. This allows us to engineer a horn-like CPW transformer with characteristic impedance varying from 50~$\Omega$ to 18~$\Omega$ (Fig.~\ref{fig:device}(b)).
The taper is designed by fixing its gap to 3 $\mathrm{\mu m}$ and varying the center-trace width from 7.8 $\mathrm{\mu m}$ to 218 $\mathrm{\mu m}$. The single CPW structure can be simulated using simple RF design suites such as TX-Line~\cite{simons2004coplanar}, and its dimensions are compatible with single-layer photo-lithography and reactive-ion etching techniques. As a result, this approach significantly reduces the complexity overhead in both design and fabrication, which is reflected by the simple construction of our device shown in Fig.~\ref{fig:device}(c).

%% file: sections/4_performance.tex
\vspace{1em}
\noindent\textbf{Device characterization}

\noindent The CIMPA is wire-bonded to a printed circuit board (PCB) and packaged using a rectangular copper box. The device is then attached to the mixing chamber stage of a dilution fridge operating at 20 mK. It is connected to room-temperature microwave apparatus as shown in Fig. \ref{fig:device}(d). A transmon qubit is integrated into the measurement setup to calibrate the signal power and detect possible back-action from the parametric process. Two QuinStar 4-8~GHz cryogenic circulators are used to route the signals, and the signals leaving the transmon are amplified by CIMPA and a high-electron-mobility-transistor amplifier (HEMT).

\begin{figure}[t!]
    \centering
    \includegraphics[width=0.47\textwidth]{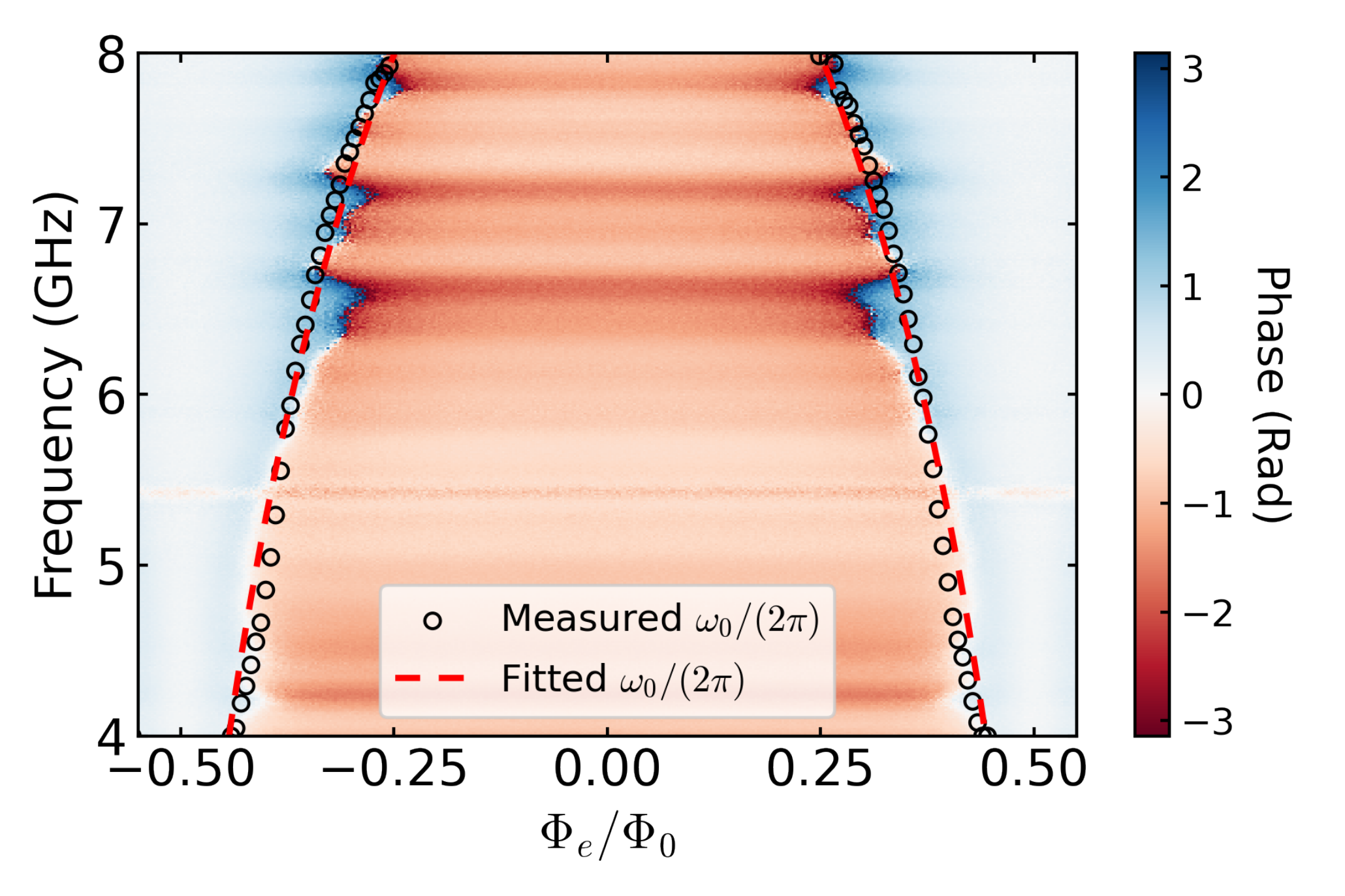}
    \caption{CIMPA's spectrum. Reflectometry spectroscopy performed using the VNA shows phase twists across the SQUID's resonant frequencies. Repeating the measurement at various flux biases yields the CIMPA's characteristic frequency spectrum, which fits the relation given by
    Eq.~\ref{eq:tune} with excellent agreement.}
    \label{fig:tune}
\end{figure}

In the first step, we extract the parameters of the nonlinear resonator by measuring the small-signal linear response of the device using a vector network analyzer (VNA). As the probe signal goes through $\omega_0/(2\pi)$, the reflected signal phase is twisted, indicating a resonant condition. When we adjust the external flux bias $\Phi_e$ of the SQUID, the resonant frequency changes according to
\begin{figure*}[t!]
    \centering
    \includegraphics[width=0.95\textwidth]{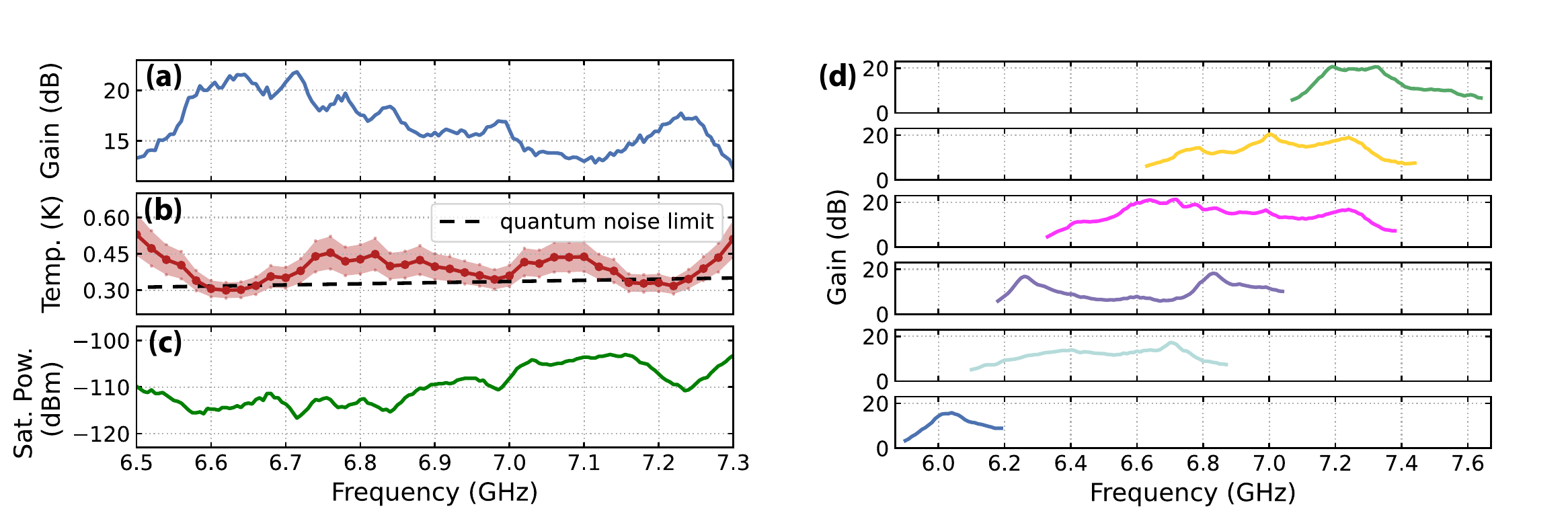}
    \caption{CIMPA's performance. (a) Instantaneous signal gain, (b) noise temperature, and (c) saturation input power of the CIMPA operated at $\Phi_e/\Phi_o=0.3$ and $f_p=13.41~\mathrm{GHz}$. (d) Signal gain at different CIMPA frequencies tuned by varying $\Phi_e$.}
    \label{fig:gain}
\end{figure*}
\begin{equation}
    \omega_0 = \left[C (L_\text{J,eff}\left|\cos(\pi\Phi_e/\Phi_0)\right|^{-1} + L_\text{geo})\right]^{-\nicefrac{1}{2}}
    \label{eq:tune},
\end{equation}
where $L_\text{geo}$ is the parasitic geometric inductance originating from the leads, $L_\text{eff}$ is the effective inductance of the SQUID, $\Phi_0$ is the magnetic flux quantum. Measuring the phase response at each flux point gives us the spectrum shown in Fig.~\ref{fig:tune}. By fitting the measured spectrum, we extract the following parameters: $L_\text{J,eff} = 69~\text{pH},~C = 4~\text{pF}$, and $L_\text{geo} < 1~\text{fH}$. The spectrum also allows us to estimate the asymmetry of the SQUID to be less than 1\%. 

We proceed to characterize the CIMPA's gain profile, noise performance, and dynamic range. We set the external flux bias to $\Phi_e=0.3\Phi_0$, and apply a microwave pump tone at frequency $\omega_p/(2\pi) = 13.41~\text{GHz}$ with 3.41~dBm power outside the fridge. we measure the microwave response using the VNA, observing over 20 (15) dB gain over a 200 (700)-\text{MHz} bandwidth (Fig.~\ref{fig:gain}(a)). Such a high increase in the bandwidth cannot be explained solely by increasing the resonator linewidth~\cite{mutus2014strong}. As discussed in Ref.~\cite{mutus2014strong}, the weak frequency dependence of the environment's characteristic impedance strongly assists the enhancement of the bandwidth, but hinders the comparison between the theory and experiments without the careful calibration of the environment's characteristic impedance. 

The noise performance of the device can be characterized via the signal-to-noise ratio (SNR) improvement $\text{SNR}_\text{imprv}$ measured by a spectrum analyzer (SA)~\cite{mutus2014strong,macklin2015near,roy2015broadband,aumentado2020superconducting}. In an amplification chain consisting of a CIMPA and a HEMT, the SNR improvement due to the CIMPA is given as 
 \begin{equation}
     \text{SNR}_\text{imprv} = \left(\frac{T_\text{CIMPA}}{T_\text{HEMT}} + \frac{1}{G}\right)^{-1},
     \label{eq:noise}
 \end{equation}
 where $T_\text{CIMPA}$ and $T_\text{HEMT}$ indicate the noise temperatures of the CIMPA and the HEMT, respectively. The extracted noise temperature with $T_\text{HEMT}=2.3$~K - $2.9$~K is shown in Fig. \ref{fig:gain}(b), which approaches the quantum noise limit. The uncertainty here originates from the estimated $T_\text{HEMT}$ range. According to the convention used in Ref.~\cite{mutus2014strong}, we define the bandwidth of our device to be 700 MHz as the frequency range over which the device approaches the quantum noise limit. The noise increase beyond this range is mainly because the CIMPA's gain is not large enough to compensate for the noise at the HEMT.

Presently, we characterize the saturation power of the device as follows. First, the relation between the actual signal power going into the CIMPA and the power set on the VNA must be established. By measuring the ac Stark shift and measurement-induced dephasing of the transmon qubit at various VNA powers, we can calibrate the cable attenuation~\cite{macklin2015near}. Then, we gradually increase the power until the gain falls off by 1 dB, and convert this using the established attenuation. The result in Fig.~\ref{fig:gain}(c) shows an average 1-dB compression point of approximately -110~dBm. Compared with the saturation power of simple JPAs coupled to a $50~\Omega$ environment (with $Q\approx15$), which is less than -120 dBm~\cite{yamamoto2008flux,mutus2014strong, ranzani2022wideband}, our saturation power is increased by around $10$ dB. This can be explained by the reduction in the environment's characteristic impedance in our design (with $Q\approx4$), as $P_\mathrm{sat}\sim C/Z_0$~\cite{mutus2014strong}. 

In addition to the instantaneous bandwidth, we can flux-tune the device and operate it at other frequencies. To showcase this capability, we calibrate the CIMPA's parameters at five different external flux points corresponding to a frequency range of 1.4 GHz, then characterize the gain performance of the device across this broad range. Figure~\ref{fig:gain}~(d) shows that the combination of the engineered broadband impedance transformer and the inherent flux-tunability allows the device to amplify signals with frequency ranging from 6~GHz to 7.4~GHz and at least 15 dB of maximum gain.

\begin{figure}[t!]
    \centering
    \includegraphics[width=0.49\textwidth]{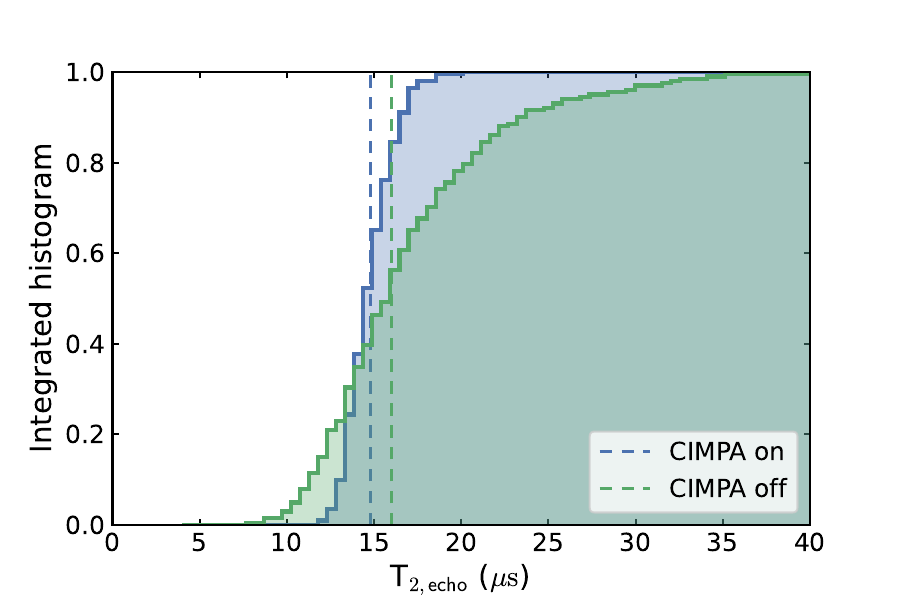}
    \caption{Transmon's coherence time statistics for CIMPA on (pumped) and off (unpumped), with the mean values equal to 14.78~$\mathrm{\mu s}$ and 16~$\mathrm{\mu s}$, and the standard deviations equal to 1.41~$\mathrm{\mu s}$ and 5.45~$\mathrm{\mu s}$, respectively. The statistics consist of 201 individual measurements in each setting.}
    \label{fig:T2}
\end{figure}

Importantly, possible back-action from the parametric process can undermine the integration of an amplifier into a qubit measurement setup. For example, spurious microwave radiation from the CIMPA can populate the readout resonator connected to the qubit, leading to photon-noise dephasing~\cite{gambetta2006qubit,kindel2016generation}. To investigate the possible back-action of the CIMPA prototype, we measure the echo time $T_{2,\text{echo}}$ of the qubit with and without amplification. To turn off the CIMPA, we remove the pump tone and flux-tune its frequency away from the qubit's readout resonator, such that the amplifier acts as a passive mirror. When the CIMPA is turned on, it provides a gain of 21 dB for the readout signal frequency at 6.633~GHz. The $T_{2,\text{echo}}$ measurement is repeated continuously over 201 iterations in each setting to provide reliable statistics despite temporal fluctuations. We attribute the wider fluctuation range associated with the measurement when the CIMPA is off to the approximately ten times longer data acquisition time. As shown in Fig.~\ref{fig:T2}, there is no significant degradation of the qubit's coherence time beyond random fluctuation when the CIMPA is operated.
\\

%% file: sections/6_summary.tex
\noindent\textbf{Discussion and outlook}

\noindent In conclusion, we show that the CIMPA's design is as simple as the JPA's, yet its performance is comparable to earlier IMPA implementations. Therefore, it fills the technological gap between JPAs and TWPAs, with potential applications ranging from qubit readout to axion dark matter detection. Compared to typical TWPAs, the prototype device reported here has a narrower bandwidth and lower dynamic range. However, we note that, on one hand, a majority of quantum experiments do not require a 3-GHz-bandwidth, and on the other hand, we can integrate SQUID array~\cite{white2023readout} or superconducting nonlinear asymmetric inductive elements (SNAIL)~\cite{ezenkova2022broadband} into the CIMPA design to increase its power handling capability.
\\

%% file: sections/author_declarations.tex
\noindent\textbf{Competing interest}

\noindent X.L., H.R., and J.L. are Anyon Technologies' employees.
\\

\noindent\textbf{Author contributions}

\noindent J.L. and B.Q. conceived and organized the project. X.L., B.Q., and H.R. fabricated the device. B.Q., L.B.N., and X.L. performed the measurement and acquired the data. B.Q. analyzed the results. N.G. assisted with the measurement apparatus. A. H. and Z.P. assisted with the cryogenic setup. W.P.L and T.C. assisted with initiating the project. H.R., J.L., D.I.S., and I.F. supervised the effort. B.Q. and L.B.N. wrote the manuscript with input from all authors.
\\